\documentclass[conference]{IEEEtran}
\IEEEoverridecommandlockouts

\usepackage{lscape}

\usepackage{multirow}     %
\usepackage{booktabs}     %
\usepackage{amsmath,amssymb,amsfonts}
\usepackage{rotating}
\usepackage{graphicx}
\usepackage{algorithm}
\usepackage{algorithmic}
\usepackage{adjustbox}
\usepackage{tabularray}
\usepackage[graphicx]{realboxes}
\usepackage{diagbox}

\def\BibTeX{{\rm B\kern-.05em{\sc i\kern-.025em b}\kern-.08em
    T\kern-.1667em\lower.7ex\hbox{E}\kern-.125emX}}
\begin{document}

\bstctlcite{IEEEexample:BSTcontrol}
\title{Hardware Efficient Accelerator for Spiking Transformer With Reconfigurable Parallel Time Step Computing\\
}

\author{\IEEEauthorblockN{Bo-Yu Chen and Tian-Sheuan Chang}
\IEEEauthorblockA{\textit{Institute of Electronics, National Yang Ming Chiao Tung University,}
Hsinchu, Taiwan
}
\IEEEauthorblockA{Email: \{brent0503.ee11, tschang\}@nycu.edu.tw}
}

\maketitle

\begin{abstract}%

This paper introduces the first low-power hardware accelerator for Spiking Transformers, an emerging alternative to traditional artificial neural networks. By modifying the base Spikformer model to use IAND instead of residual addition, the model exclusively utilizes spike computation. The hardware employs a fully parallel tick-batching dataflow and a time-step reconfigurable neuron architecture, addressing the delay and power challenges of multi-timestep processing in spiking neural networks. This approach processes outputs from all time steps in parallel, reducing computation delay and eliminating membrane memory, thereby lowering energy consumption. The accelerator supports 3x3 and 1x1 convolutions and matrix operations through vectorized processing, meeting model requirements. Implemented in TSMC's 28nm process, it achieves 3.456 TSOPS (tera spike operations per second) with a power efficiency of 38.334 TSOPS/W at 500MHz, using 198.46K logic gates and 139.25KB of SRAM.

~
\end{abstract}

\section{Introduction}
Spiking neural networks (SNNs) have become popular recently as an alternative to artificial neural networks (ANNs) due to its binary spike operation and high sparsity to solve the high energy computation and memory bandwidth in ANNs. An SNN operation in a time step integrates its incoming spikes, fires a spike if the summation of input accumulation and current membrane potential exceeds the threshold or keeps it to the membrane potential if smaller. These SNNs have been successfully applied to convolutional neural networks (CNN)~\cite{diehl2015unsupervised, rathi2020enabling} as well as transformers~\cite{hu2021spiking, li2022spikeformer, zhou2022spikformer, zhou2024spikformer}, which can achieve competitive performance to ANN~\cite{zhou2024spikformer}.

For studies to apply spiking neural networks (SNNs) to Transformer, Li et al.~\cite{li2022spikeformer} proposed the first work that combines Transformer with SNNs. However, the sequence pooling in their proposed architecture involves a significant amount of floating-point multiplication, division, and exponential operations in the Softmax computation, and thus the algorithm still relies on the traditional attention mechanism. Zhou et al.~\cite{zhou2022spikformer, zhou2024spikformer} proposed Spikformer, which combines SNNs with self-attention. By using binary-spike-forms of Q, K, and V for matrix operations, Spikformer avoids negative values and eliminates Softmax to reduce computational complexity, as shown in the Fig.~\ref{Non-spike computation of Spikformer}. However, Spikformer only reduces the computational complexity of self-attention. It still needs non-spike calculation due to the residual summation.

Beyond above model developments, hardware accelerators for SNNs also become popular in providing low-cost and low-power implementations. However, current designs only consider CNN models~\cite{lee2020reconfigurable, chen2024sibrain}, and none have considered transformer models. Compared to ANN accelerators, due to the additional time steps in spiking neural networks (SNNs), accelerators for SNNs require more time for inference compared to traditional ANN accelerators. Thus, SpinalFlow~\cite{narayanan2020spinalflow} has proposed a serial tick-batching dataflow that processes different time steps of the same layer successively and then changes to other layers to reduce the buffer size of the membrane. However, this serial approach needs to access weight and membrane SRAM repeatedly for different time steps.  To solve this problem, accelerators ~\cite{lee2022parallel, lee2020reconfigurable} uses systolic arrays to compute SNNs simultaneously in both spatial and temporal domains. However, they require additional buffers to temporarily store and transfer values such as inputs, weights, and membrane potentials. Furthermore, each time step can only be computed after receiving information from the previous time step, making it impossible to process all time steps in parallel and incurring long latency. Although Sibrain~\cite{chen2024sibrain} can parallelize the calculation of membrane values at each time step, it still requires the sequential transmission of membrane information from the previous time step to the next in order to generate the output spikes.

Addressing the above issues, this paper presents a low-power hardware accelerator for spiking vision transformers. For the model, we propose an all-spike computation spiking transformer, Spike-IAND-Former, by replacing residual additions with IAND operations to reduce area cost and power consumption. For the hardware, to solve the extra latency and power due to multiple time steps, we propose a fully parallel tick-batching flow instead of a serial one to process different time steps at the same time, which can be reconfigured to support different parallelism as well. The hardware architecture adopts a vectorized processing dataflow to support different computational patterns like convolution and matrix multiplication in the model. The final implementation can achieve 38.334 TSOPS/W power efficiency when operated at 500MHz, exceeding existing designs.

The rest of the paper is organized as follows. Section II shows the modified spiking transformer. Section III presents the proposed hardware design. Section IV shows the experimental results. Finally, this paper is concluded in Section V.

\section{Hardware Friendly Spiking Vision Transformer}
\begin{figure}[htbp]
\centering
\includegraphics[height=!,width=1.0\linewidth,keepaspectratio=true]{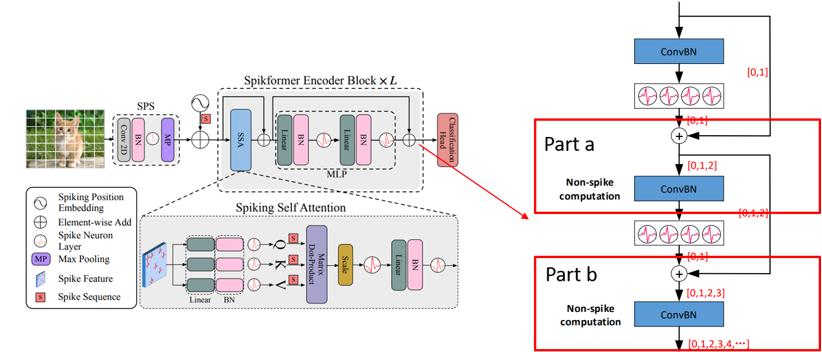}
\caption{Spikformer~\cite{zhou2022spikformer} and its non-spike computation part.}
\label{Non-spike computation of Spikformer}
\end{figure}

The adopted model is based on Spikformer~\cite{zhou2022spikformer, zhou2024spikformer} as shown in Fig.~\ref{Non-spike computation of Spikformer} that effectively addresses the non-spike computation issue in self-attention and achieves comparable performance to ANN. However, the residual summation in this model causes non-spike computations (values are no longer 0/1) in the convolution layers due to the addition operations. This not only limits the energy efficiency advantages of SNN Transformers but also complicates their deployment and optimization on SNN hardware instead of simple AND gates for multiplication.

To solve this non-spike computation, we adopt the element-wise-IAND~\cite{fang2021deep} as the operator in the residual block to address the non-spike computation issues that the model may encounter. The IAND operation is $x*(1-ConvBN(x))$, where x is the input, and ConvBN is convolution and BN operations as in Fig.~\ref{Non-spike computation of Spikformer}. Since both x and ConvBN(x) are spike, their multiplication can be simplified as an AND operation.

\begin{figure}[htbp]
\centering
\includegraphics[height=!,width=1.0\linewidth,keepaspectratio=true]{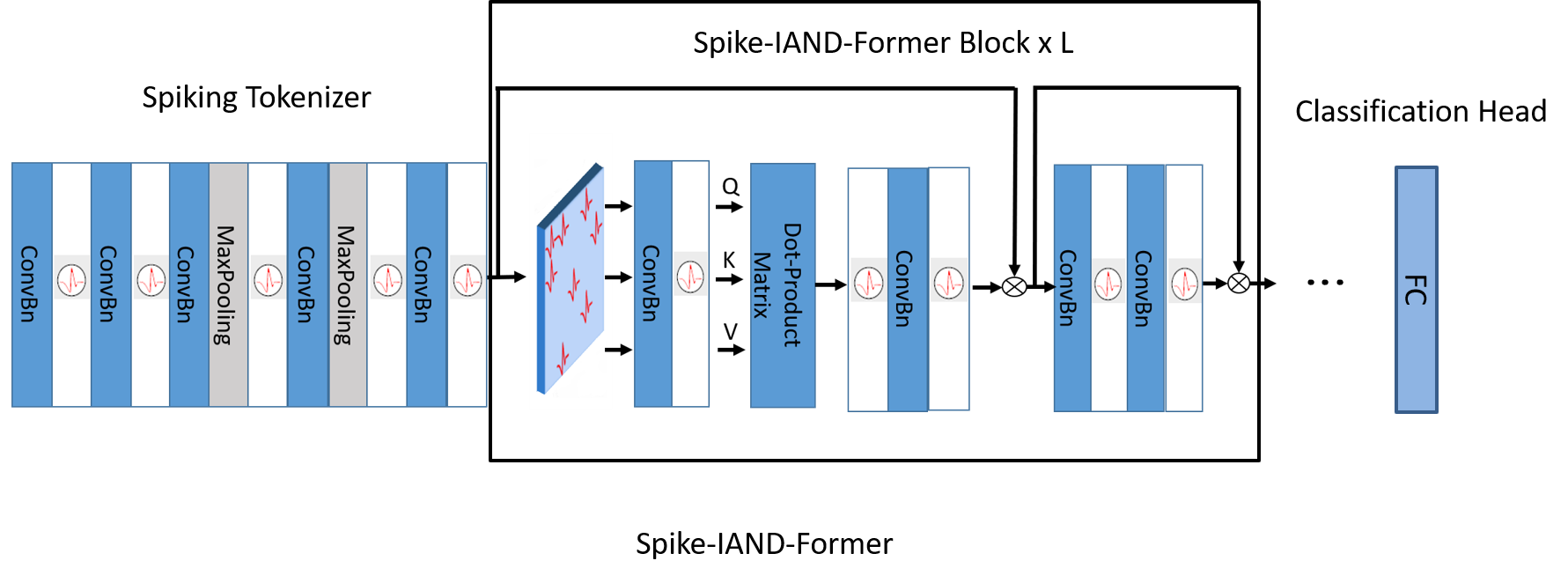}
\caption{Spike-IAND-Former}
\label{Spike-IAND-Former}
\end{figure}

The proposed model is shown in the Fig.~\ref{Spike-IAND-Former}, denoted as Spike-IAND-Former, with leaky integrate-and-fire (LIF) neurons.  The threshold of LIF neurons is set to 0.5, and the leakage term is set to 0.25. The model mainly includes the Spiking Tokenizer, Spike-IAND-Former Block, and Classification Head. The Spiking Tokenizer is responsible for generating convolutional spiking patch embeddings and downsampling the image size. In the Spiking Tokenizer, the first convolutional layer is designated as the encoding layer~\cite{wu2019direct}, which is responsible for converting the 8-bit image input signals into spike signals within the time steps. In other words, it transforms the high-precision data in the spatial domain into spike signals in the temporal domain. The Spike-IAND-Former Block functions similarly to the Transformer residual block in a traditional ViT, but we chose to use element-wise-IAND~\cite{fang2021deep} as the operator in the residual block to address the non-spike computation issues that the model may encounter. The Classification Head is used for classifying the final results of the model. This model can achieve competitive performance to ANN with time steps equal to 4 and have slightly lower performance for a single time step, as shown in the result section. Thus, our design will consider to support up to four time steps.

\section{Proposed Hardware 
}
\label{chapter:Reconfigurable Spatial-Temporal Parallel Spiking Neural Network Accelerator for Vision Transformer 
}

\subsection{Overview}

\begin{figure}[htbp]
\centering
\includegraphics[height=!,width=1.0\linewidth,keepaspectratio=true]{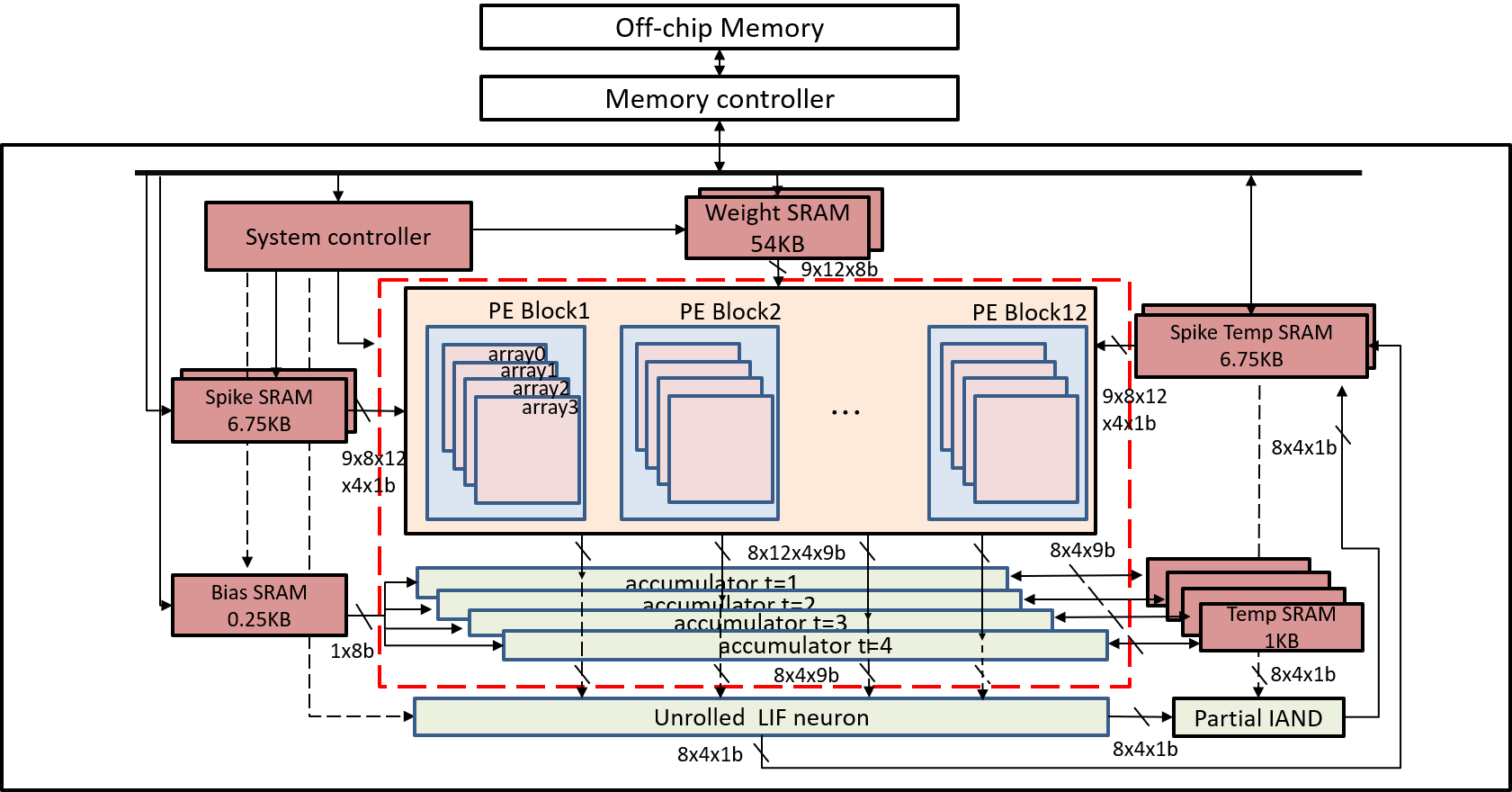}
\caption{The proposed system architecture}
\label{The proposed system architecture}
\end{figure}

Fig.~\ref{The proposed system architecture} illustrates the proposed system architecture. The system accesses weights and inputs from off-chip memory through the memory controller and stores them into weight SRAMs and spike SRAMs. Each PE block can process four time steps of an input spike channel in parallel, while each PE array is responsible for computing the operations for each time step on the input channel to complete 3x3 convolution, 1x1 convolution, and matrix multiplications. A total of 12 PE blocks can simultaneously process four time steps on 12 input channels. The accumulator is responsible for accumulating the outputs of the four time steps across the 12 channels and storing the results in temp SRAMs, then accumulating these with the results already stored in temp SRAMs to produce the output of one output channel. The unrolled LIF neurons convert the output of the output channel into output spikes and can generate output for four time steps simultaneously. The results are saved in  spike temp SRAMs and then transferred back to off-chip memory through the memory controller. The 8-bit image input is split into bitplanes for computations to reuse the spike input PE blocks.

The proposed design adopts the proposed fully parallel tick-batching data flow. A typical data flow in ANN accelerators will use layer-by-layer processing to execute a model. However, this is not suitable for SNN due to multiple time step in a layer execution. Thus, ~\cite{narayanan2020spinalflow} proposed the tick-batching data flow that processes all time steps in a layer before proceeding to the next layer. But the flow in ~\cite{narayanan2020spinalflow} processes time step one-by-one, which will need to access weight buffers repeatedly for different time steps and needs extra SRAM buffers for membrane potential. These extra accesses consume significant power and easily offset the benefits of spiking computation~\cite{lien2021vsa}. Thus, we propose a fully parallel tick-batching data flow. In the SNN operations, the multiplication between input and weight has no data dependency between time steps. This allows parallel processing of input for all time steps. However, the neuron output of each time step depends on the membrane potential of previous time steps. To resolve this data dependency, we unroll the LIF neuron loop for all time steps. Specifically, for the Spiking-IAND-former model with four time steps, we will use four times parallelism in our PE block and unrolled LIF neurons for four time-step inputs. This approach brings the benefit of lower latency and lower power with area increase. The area increase due to this parallelism has a small overhead due to spike operations as shown in the result section. However, this can reduce the required latency by four. Besides, the buffer of the membrane potential is eliminated, which occupies more area than the area overhead due to multi-bit values. The access to the weight SRAM is also reduced to once instead of four since all time steps can share the same weight.  

\subsection{PE Blocks}

\begin{figure}[tbp]
\centering
\includegraphics[height=!,width=1.0\linewidth,keepaspectratio=true]{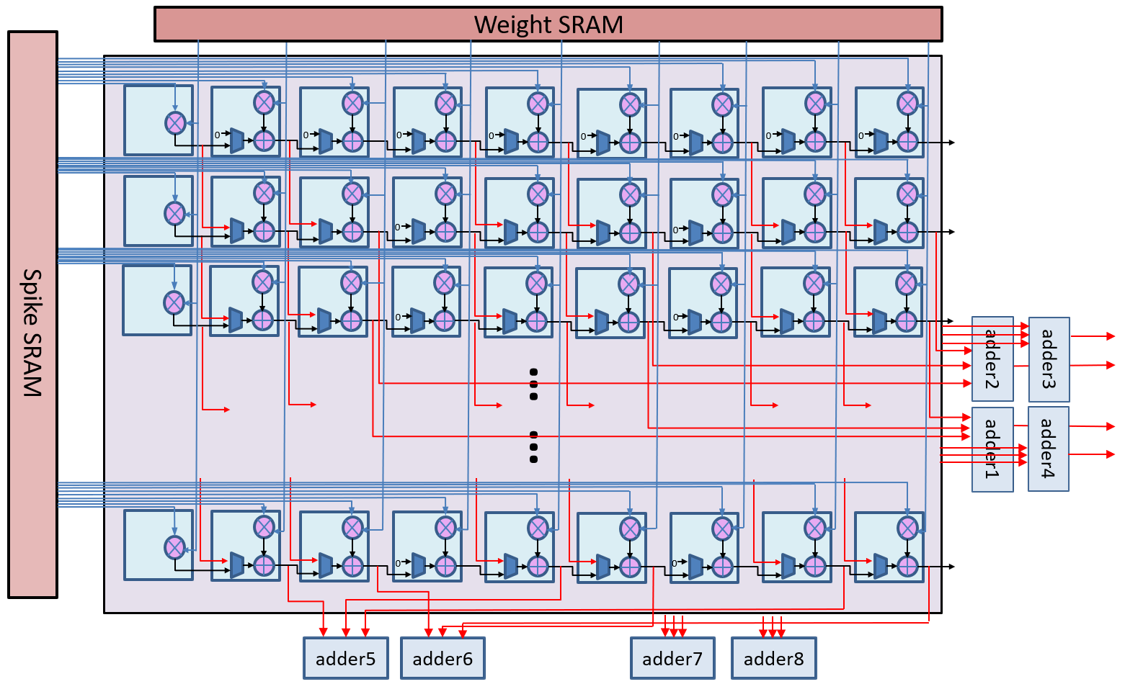}
\caption{The proposed PE array}
\label{The proposed PE array}
\end{figure}

Each PE block is composed of four PE arrays, and each PE array consists of 8x9 PEs. As shown in Fig.~\ref{The proposed PE array}, eight input spike signals broadcast horizontally, and nine weights broadcast vertically. When performing the 3x3 convolution, the data is accumulated in the direction indicated by the red arrows in Fig.~\ref{The proposed PE array}, and the result is sent to the accumulator to complete the summation of 12 channels. For the 1x1 convolution and matrix multiplication, the data is accumulated in the direction indicated by the black arrows in Fig.~\ref{The proposed PE array}.

\subsection{Reconfigurable Unrolled LIF Neuron}

\begin{figure}[tbp]
\centering
\includegraphics[height=!,width=1.0\linewidth,keepaspectratio=true]{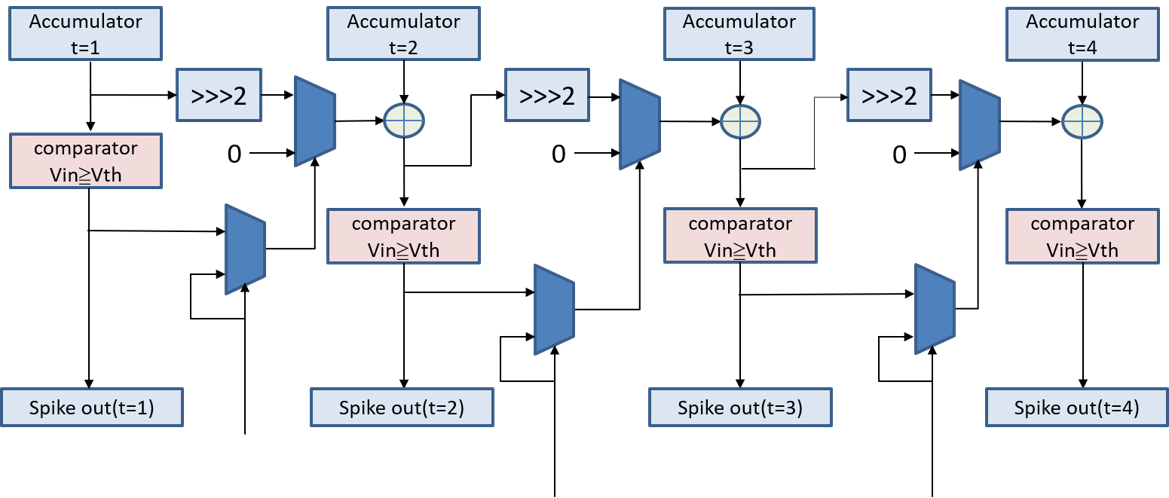}
\caption{The proposed reconfigurable unrolled LIF neuron. The MUX selector input (from left to right) will be set to 111/101/000 for the time step=4/2/1, respectively. }
\label{The proposed reconfigurable unrolled LIF neuron}
\end{figure}

Fig.~\ref{The proposed reconfigurable unrolled LIF neuron} shows the proposed unrolled LIF neuron architecture. By unrolling the LIF neuron to simultaneously compute four time steps, we achieve spatial-temporal parallel acceleration. This approach of parallel processing outputs of all time steps reduces hardware computation delay while also avoiding the need to access membrane memory. As a result, the inference time of our SNN accelerator can be as short as that of an ANN accelerator. This neuron can be configured to support different time steps with three multiplexers to control how to propagate data to neighbor neurons for different time steps. 

\begin{figure}[tbp]
\centering
\includegraphics[height=!,width=1.0\linewidth,keepaspectratio=true]{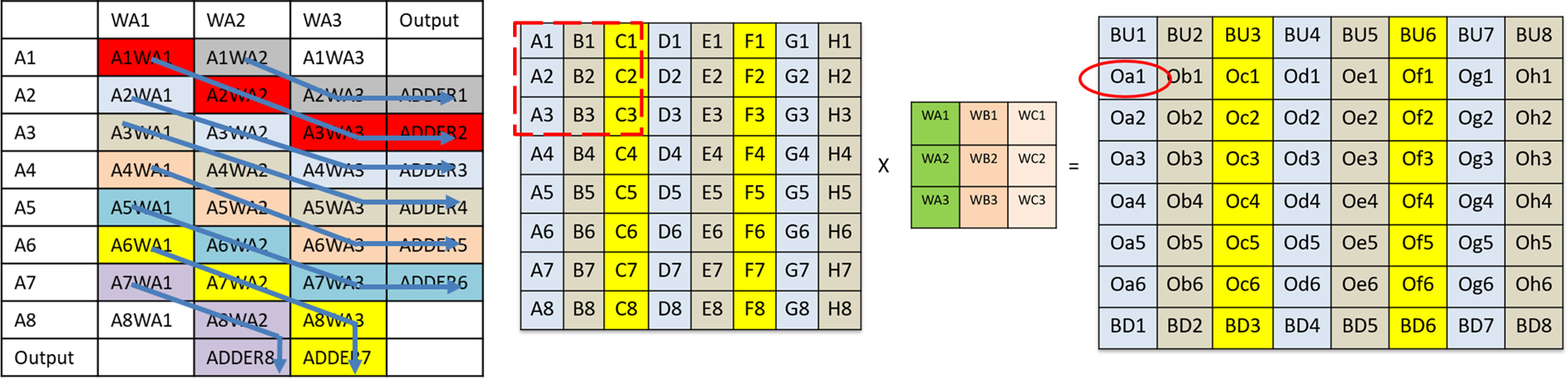}
\caption{Data flow of 3x3 convolution}
\label{A convolution example with 8×8 inputs, 3×3 weights, and 8×8 outputs}
\end{figure}

\begin{table}[tbp]
\centering
\caption{Result of different models on ImageNet, A-B (e.g. 8-384) denotes layer numbers and embeddeding dimensions. 400E denotes 400 epoches.}
\begin{tblr}{
  cells = {c},
  hline{1-3,6-8} = {-}{},
}
Methods                     & Architecture                                                                                                                   & Time step                   & Accuracy                                \\
ANN~\cite{zhou2022spikformer}                     & Trans.-8-512                                                                                                              & 1                           & 80.80                                       \\
Hybrid training~\cite{rathi2020enabling}         & ResNet-34                                                                                                                      & 250                         & 61.48                                       \\
Spiking ResNet~\cite{hu2021spiking}          & ResNet-34                                                                                                                      & 350                         & 71.61                                       \\
STBP-tdBN~\cite{zheng2021going}                & Spiking-ResNet-34                                                                                                              & 6                           & 63.72                                       \\
Spikformer~\cite{zhou2022spikformer}              & {
  8-384
  \\8-512
  \\8 768
  }                                                             & {
  4
  \\4
  \\4
  }       & {
  70.24
  \\73.38
  \\74.81
  }           \\
\textbf{Spike-IAND-former } & {
  8-384
  \\8-512
  \\8-768
  \\\textbf{8-768 400E}} & {
  4
  \\4
  \\4
  \\4
  } & {
  70.32
  \\73.45
  \\74.89
  \\74.97
  } 
\end{tblr}
\label{Result of different models on ImageNet}
\end{table}

\begin{table*}[htbp]
\centering
\caption{Comparison with other designs. Normalized area and power are scaled to 28nm. }
\begin{tabular}{|l|c|c|c|c|c|c|c|} 
\hline
                                                                                               & \textbf{This work} & SpinalFlow~\cite{narayanan2020spinalflow} & TCAS II'21~\cite{chen202167} & ~\cite{xie2023energy} & VSA~\cite{lien2021vsa}      & ~\cite{kang202324} & BW-SNN~\cite{chuang202090nm}   \\ 
\hline
Technology                                                                                     & 28nm               & 28nm       & 28nm       & 28nm      & 40nm     & 65nm       & 90nm     \\ 
\hline
Voltage (V)                                                                                    & 0.9                & -          & 0.81       & 0.9       & 0.9      & 0.9        & 0.6      \\ 
\hline
Frequency (MHz)                                                                                & 500                & 200        & 500        & 650       & 500      & 250        & 10       \\ 
\hline
Network architecture in SNN                                                                    & Transformer        & CNN        & CNN        & CNN       & CNN      & CNN        & CNN      \\ 
\hline
Reconfigurable                                                                                 & Yes                & Yes        & Yes        & Yes       & Yes      & Yes        & No       \\ 
\hline
Spatial-Temporal Processing                                                                    & Yes                & No         & No         & Yes       & No       & No         & No       \\ 
\hline
Weight Precision (bits)                                                                        & 8                  & 8          & 8          & 8         & binary   & 8          & binary   \\ 
\hline
PE number                                                                                      & 3456               & 128        & -          & 1148      & 2304     & -          & 8208     \\ 
\hline
SRAM (KB)                                                                                      & 139.25             & 585        & 240        & 143       & 230.31   & 384        & 12.75    \\ 
\hline
Peak Throughput (GSOPS)                                                                        & 3456               & 51.2       & 1150       & 746       & 2304     & 1020       & 64.46    \\ 
\hline
Area (KGE) (logic only)                                                                        & 198.46             & -          & 54.72      & -         & 114.98   & -          & 225      \\ 
\hline
Area eff. (GSOPS/KGE)                                                                          & 17.414             & -          & 21.016     & -         & 20.038   & -          & 0.286    \\ 
\hline
Normalized area efficiency (GSOPS/KGE)                                         & 17.414             & -          & 21.016     & -         & 28.625   & -          & 0.92     \\ 
\hline
Core power (mW)                                                                                & 90.153             & 162.4      & 149.3      & 20.22     & 88.968   & 181.6      & 0.625    \\ 
\hline
Power eff. (TSOPS/W)                                                                           & 38.334             & 0.315      & 7.702      & 36.89     & 25.9     & 5.616      & 103.14   \\ 
\hline
Normalized power efficiency (TSOPS/W)                                          & \textbf{38.334}    & -          & 6.238      & 36.89     & 37       & 13.037     & 147.342  \\ 
\hline
\end{tabular}
\label{Comparison with other designs}
\end{table*}

\subsection{Data Flow}
\subsubsection{Data Flow of 3x3 Convolution}
Fig.~\ref{A convolution example with 8×8 inputs, 3×3 weights, and 8×8 outputs} shows the data flow of PE arrays for 3x3 convolutions. As shown in the figure, we divide the 8x9 PE array into three 8x3 PE arrays to facilitate understanding the operation of the 3x3 convolution, where the input signals are broadcasting horizontally, the weights are broadcasting vertically and the partial sum is accumulated diagonally. With this, each clock cycle can produce eight elements of the output channel.

\subsubsection{Data Flow of 1x1 Convolution and matrix multiplication}
For the 1x1 convolution data flow, it broadcasts the inputs horizontally, and the weights vertically and accumulates the partial sum horizontally. Note that the each input is different, but this is acceptable due to simple spike input. The 8x9 PE array is capable of performing 1x1 convolution on one column of 9 different 8x8 input channels simultaneously. In our data flow, it takes eight cycles to complete the 1x1 convolution for the 9 different 8x8 input channels. The matrix multiplication uses the same flow as this one.

\section{Experimental Result}
\subsection{Simulation Results of the Proposed Model}
The proposed model has been implemented by Pytorch. The model training adopts the AdamW optimizer with the batch size set to 256, and the cosine annealing learning rate starting from $5 \times 10^{-4}$ for 400 epochs.

Table~\ref{Result of different models on ImageNet} presents a comparison between our model and other models~\cite{rathi2020enabling,hu2021spiking,zheng2021going,zhou2022spikformer} on the ImageNet dataset. From the table, we can see that our model achieves more accurate results compared to other state-of-the-art models, even with the same or fewer time steps. The accuracy of CIFAR-10 is 95.69 for the same configuration of the model. Based on this, the required time step can be decreased to 2 or 1 with a progressive time step reduction~\cite{chowdhury2021one} to 92.93 and 91.34, respectively.

\subsection{Hardware Implementation and Comparisons}
 The proposed design was developed using Verilog and synthesized with the Synopsys Design Compiler using the TSMC 28nm CMOS technology. Power consumption was measured using Synopsys PrimeTime PX. The proposed design achieves a peak performance of 3456GSOPS (giga spike operations per second) with 198.46K gate count and 139.25KB SRAM and 90.153mW when operated at the 500MHz clock frequency. When tested on CIFAR-10 with a time step of 4, it achieves a processing capability of 46.72 frames per second. The lower power is contributed by the high sparsity of the model (73.88\% zeros in average for the activation map) and our low power design. In particular, using unrolled LIF reduces weights access amounts from weight SRAMs by 43.2\%. This also eliminates the need and the power of the membrane SRAM. Thus, the memory only accounts for 43\% of the total power, while the remaining logic operations account for 57\% of the total power.

To the best of our knowledge, we are the first to develop an SNN accelerator for Vision Transformers.  Table~\ref{Comparison with other designs} shows the performance of our hardware and a comparison with other SNN designs. Compared to the other design~\cite{xie2023energy} that support spatial-temporal parallel processing, the use of unrolled LIF reduces weights access amounts from SRAMs by 43.2\% while also eliminating the need for accessing membrane SRAM, resulting in better power efficiency. ~\cite{chuang202090nm} is a five-layer SNN ASIC that is specifically designed to reduce power consumption by minimizing memory accesses. Nevertheless, that design has a very low area efficiency and cannot be used for different models.

\section{Conclusion}
This paper proposes a hardware-efficient spiking transformer accelerator.  The proposed design solves the non-spike computation problem in spiking transformers by replacing residual addition with element-wise-IAND to make the whole model spike I/O only. The diverse layer types, 3x3 convolution, 1x1 convolution, and matrix multiplication, in this vision spiking transformer are supported by a reconfigurable vectorwise data flow. The long computation delay due to multiple time steps in SNN is solved by the proposed reconfigurable LIF architecture for spatial-temporal parallel processing to reduce delay, avoid the need of membrane memory and thus reduce power consumption. The design implemented in a TSMC 28nm CMOS process only requires 198.46K logic gates and 139.25KB of SRAM to achieve a throughput of 3456GSOPS (giga spike operations per second). Its power efficiency reaches up to 38.334TSOPS/W (tera spike operations per second/W) when operating at a frequency of 500MHz.

\bibliographystyle{IEEEtran}
\bibliography{bib/ieeeBSTcontrol,bib/thesis}

\begin{thebibliography}{10}
\providecommand{\url}[1]{#1}
\csname url@samestyle\endcsname
\providecommand{\newblock}{\relax}
\providecommand{\bibinfo}[2]{#2}
\providecommand{\BIBentrySTDinterwordspacing}{\spaceskip=0pt\relax}
\providecommand{\BIBentryALTinterwordstretchfactor}{4}
\providecommand{\BIBentryALTinterwordspacing}{\spaceskip=\fontdimen2\font plus
\BIBentryALTinterwordstretchfactor\fontdimen3\font minus \fontdimen4\font\relax}
\providecommand{\BIBforeignlanguage}[2]{{%
\expandafter\ifx\csname l@#1\endcsname\relax
\typeout{** WARNING: IEEEtran.bst: No hyphenation pattern has been}%
\typeout{** loaded for the language `#1'. Using the pattern for}%
\typeout{** the default language instead.}%
\else
\language=\csname l@#1\endcsname
\fi
#2}}
\providecommand{\BIBdecl}{\relax}
\BIBdecl

\bibitem{diehl2015unsupervised}
P.~U. Diehl and M.~Cook, ``Unsupervised learning of digit recognition using spike-timing-dependent plasticity,'' \emph{Frontiers in computational neuroscience}, vol.~9, p.~99, 2015.

\bibitem{rathi2020enabling}
N.~Rathi, G.~Srinivasan, P.~Panda, and K.~Roy, ``Enabling deep spiking neural networks with hybrid conversion and spike timing dependent backpropagation,'' \emph{arXiv preprint arXiv:2005.01807}, 2020.

\bibitem{hu2021spiking}
Y.~Hu, H.~Tang, and G.~Pan, ``Spiking deep residual networks,'' \emph{IEEE Transactions on Neural Networks and Learning Systems}, vol.~34, no.~8, pp. 5200--5205, 2021.

\bibitem{li2022spikeformer}
Y.~Li, Y.~Lei, and X.~Yang, ``Spikeformer: a novel architecture for training high-performance low-latency spiking neural network,'' \emph{arXiv preprint arXiv:2211.10686}, 2022.

\bibitem{zhou2022spikformer}
Z.~Zhou, Y.~Zhu, C.~He, Y.~Wang, S.~Yan, Y.~Tian, and L.~Yuan, ``Spikformer: when spiking neural network meets transformer,'' \emph{arXiv preprint arXiv:2209.15425}, 2022.

\bibitem{zhou2024spikformer}
Z.~Zhou, K.~Che, W.~Fang, K.~Tian, Y.~Zhu, S.~Yan, Y.~Tian, and L.~Yuan, ``Spikformer v2: Join the high accuracy club on {ImageNet} with an {SNN} ticket,'' \emph{arXiv preprint arXiv:2401.02020}, 2024.

\bibitem{lee2020reconfigurable}
J.-J. Lee and P.~Li, ``Reconfigurable dataflow optimization for spatiotemporal spiking neural computation on systolic array accelerators,'' in \emph{2020 IEEE 38th International Conference on Computer Design (ICCD)}.\hskip 1em plus 0.5em minus 0.4em\relax IEEE, 2020, pp. 57--64.

\bibitem{chen2024sibrain}
Y.~Chen, W.~Ye, Y.~Liu, and H.~Zhou, ``Sibrain: a sparse spatio-temporal parallel neuromorphic architecture for accelerating spiking convolution neural networks with low latency,'' \emph{IEEE Transactions on Circuits and Systems I: Regular Papers}, 2024.

\bibitem{narayanan2020spinalflow}
S.~Narayanan, K.~Taht, R.~Balasubramonian, E.~Giacomin, and P.-E. Gaillardon, ``Spinalflow: an architecture and dataflow tailored for spiking neural networks,'' in \emph{2020 ACM/IEEE 47th Annual International Symposium on Computer Architecture (ISCA)}.\hskip 1em plus 0.5em minus 0.4em\relax IEEE, 2020, pp. 349--362.

\bibitem{lee2022parallel}
J.-J. Lee, W.~Zhang, and P.~Li, ``Parallel time batching: systolic-array acceleration of sparse spiking neural computation,'' in \emph{2022 IEEE International Symposium on High-Performance Computer Architecture (HPCA)}.\hskip 1em plus 0.5em minus 0.4em\relax IEEE, 2022, pp. 317--330.

\bibitem{fang2021deep}
W.~Fang, Z.~Yu, Y.~Chen, T.~Huang, T.~Masquelier, and Y.~Tian, ``Deep residual learning in spiking neural networks,'' \emph{Advances in Neural Information Processing Systems}, vol.~34, pp. 21\,056--21\,069, 2021.

\bibitem{wu2019direct}
Y.~Wu, L.~Deng, G.~Li, J.~Zhu, Y.~Xie, and L.~Shi, ``Direct training for spiking neural networks: faster, larger, better,'' in \emph{Proceedings of the AAAI conference on artificial intelligence}, vol.~33, no.~01, 2019, pp. 1311--1318.

\bibitem{lien2021vsa}
H.-H. Lien, C.-W. Hsu, and T.-S. Chang, ``Vsa: reconfigurable vectorwise spiking neural network accelerator,'' in \emph{2021 IEEE International Symposium on Circuits and Systems (ISCAS)}.\hskip 1em plus 0.5em minus 0.4em\relax IEEE, 2021, pp. 1--5.

\bibitem{zheng2021going}
H.~Zheng, Y.~Wu, L.~Deng, Y.~Hu, and G.~Li, ``Going deeper with directly-trained larger spiking neural networks,'' in \emph{Proceedings of the AAAI conference on artificial intelligence}, vol.~35, no.~12, 2021, pp. 11\,062--11\,070.

\bibitem{chen202167}
Q.~Chen, G.~He, X.~Wang, J.~Xu, S.~Shen, H.~Chen, Y.~Fu, and L.~Li, ``A 67.5 $\mu$j/prediction accelerator for spiking neural networks in image segmentation,'' \emph{IEEE Transactions on Circuits and Systems II: Express Briefs}, vol.~69, no.~2, pp. 574--578, 2021.

\bibitem{xie2023energy}
C.~Xie, Z.~Shao, Z.~Chen, Y.~Du, and L.~Du, ``An energy-efficient spiking neural network accelerator based on spatio-temporal redundancy reduction,'' \emph{IEEE Transactions on Very Large Scale Integration (VLSI) Systems}, 2023.

\bibitem{kang202324}
L.~Kang, X.~Yang, C.~Zhang, S.~Yu, R.~Dou, W.~Li, C.~Shi, J.~Liu, N.~Wu, and L.~Liu, ``A 24.3 $\mu$j/image snn accelerator for dvs-gesture with ws-los dataflow and sparse methods,'' \emph{IEEE Transactions on Circuits and Systems II: Express Briefs}, vol.~70, no.~11, pp. 4226--4230, 2023.

\bibitem{chuang202090nm}
P.-Y. Chuang, P.-Y. Tan, C.-W. Wu, and J.-M. Lu, ``A 90nm 103.14 tops/w binary-weight spiking neural network cmos asic for real-time object classification,'' in \emph{2020 57th ACM/IEEE Design Automation Conference (DAC)}.\hskip 1em plus 0.5em minus 0.4em\relax IEEE, 2020, pp. 1--6.

\bibitem{chowdhury2021one}
S.~S. Chowdhury, N.~Rathi, and K.~Roy, ``One timestep is all you need: Training spiking neural networks with ultra low latency,'' \emph{arXiv preprint arXiv:2110.05929}, 2021.

\end{thebibliography}


\end{document}